\documentclass[conference]{IEEEtran}
\IEEEoverridecommandlockouts
\usepackage{cite}
\usepackage{amsmath,amssymb,amsfonts}
\usepackage{algorithmic}
\usepackage{graphicx}
\usepackage{textcomp}
\usepackage{xcolor}
\usepackage{tabularx}
\usepackage{booktabs}
\usepackage{multirow}
\usepackage{pgfplots}
\usepackage{graphicx}
\usepackage{subcaption}
\usepackage{caption}
\usepackage{float}
\def\BibTeX{{\rm B\kern-.05em{\sc i\kern-.025em b}\kern-.08em
    T\kern-.1667em\lower.7ex\hbox{E}\kern-.125emX}}

\makeatletter
\newcommand{\linebreakand}{%
  \end{@IEEEauthorhalign}
  \hfill\mbox{}\par
  \mbox{}\hfill\begin{@IEEEauthorhalign}
}
\makeatother

\begin{document}

\title{EEG-EMG FAConformer: Frequency Aware Conv-Transformer for the fusion of EEG and EMG 
 \\
}

\author{
\IEEEauthorblockN{ZhengXiao He*†\thanks{*These authors contributed equally to this work.}\thanks{\textsuperscript{†}Corresponding author: ZhengXiao He (1950095@tongji.edu.cn)}}
\IEEEauthorblockA{\textit{College of Electronic and Information Engineering} \\
\textit{Tongji University}\\
Shanghai, China \\
0009-0008-4147-3310}
\and
\IEEEauthorblockN{Minghong Cai*}
\IEEEauthorblockA{\textit{School of Software Engineering} \\
\textit{Tongji University}\\
Shanghai, China \\
2053769@tongji.edu.cn}
\and
\IEEEauthorblockN{Letian Li*}
\IEEEauthorblockA{\textit{School of Software Engineering} \\
\textit{Tongji University}\\
Shanghai, China \\
liletian@tongji.edu.cn}

\linebreakand 

\IEEEauthorblockN{Siyuan Tian*}
\IEEEauthorblockA{\textit{College of Electronic and Information Engineering} \\
\textit{Tongji University}\\
Shanghai, China \\
2053422@tongji.edu.cn}
\and
\IEEEauthorblockN{Ren-Jie Dai}
\IEEEauthorblockA{\textit{Department of Computer Science and Engineering} \\
\textit{Shanghai Jiao Tong University}\\
Shanghai, China \\
renjiedai01@163.com}
}

\maketitle

\begin{abstract}
Motor pattern recognition paradigms are the main forms of Brain-Computer Interfaces(BCI) aimed at motor function rehabilitation and are the most easily promoted applications. In recent years, many researchers have suggested encouraging patients to perform real motor control execution simultaneously in MI-based BCI rehabilitation training systems. Electromyography (EMG) signals are the most direct physiological signals that can assess the execution of movements. Multimodal signal fusion is practically significant for decoding motor patterns. Therefore, we introduce a multimodal motion pattern recognition algorithm for EEG and EMG signals: EEG-EMG FAConformer, a method with several attention modules correlated with temporal and frequency information for motor pattern recognition. We especially devise a frequency band attention module to encode EEG information accurately and efficiently. What's more, modules like Multi-Scale Fusion
Module, Independent Channel-Specific Convolution Module(ICSCM), and Fuse Module which can effectively eliminate irrelevant information in EEG and EMG signals and fully exploit hidden dynamics are developed and show great effects. Extensive experiments show that EEG-EMG FAConformer surpasses existing methods on Jeong2020 dataset, showcasing outstanding performance, high robustness and impressive stability.

\end{abstract}


\section{Introduction}
Our research focuses on motor pattern recognition\cite{lotze2006motor} using EEG and EMG signals. Since the 1990s, top research institutions have advanced brain-computer interface (BCI) systems. Early leaders included Graz University of Technology, University of Tübingen, and the Wadsworth Center\cite{vaughan2006wadsworth}. Graz University, led by Pfurtscheller\cite{pfurtscheller1999event}, pioneered using event-related potentials for motor imagery tasks. Since 2003, BCI competitions have furthered data processing and classification techniques.

Despite progress, processing EEG signals remains challenging due to low signal-to-noise ratio, noise, artifacts, and individual differences. Extracting critical information and improving algorithm interpretability are key challenges. Recent deep learning methods like Liu\cite{liu2020parallel}, Fan\cite{fan2021bilinear}, EEG Conformer\cite{song2022eeg}, EEG-Deformer\cite{she2023improved}, CNN-LSTM\cite{li2022motor}, EEGNet\cite{lawhern2018eegnet}, EEG-TCNet\cite{ingolfsson2020eeg}, and LMDA-Net\cite{miao2023lmda} have shown promise, outperforming traditional methods. However, these methods often overlook the correlation between EEG and EMG in motor pattern recognition. Although previous works can capture both short period and long period features in some way, they did not effectively get rid of the rebundant information in irrelevant frequency band and noises mixed in the signals. That's what we are looking for: a method that can eliminate irrelevant information and noises. Also, it can grasp temporal and frequency information in both short periods and long periods for discerning  dynamics embedded in the EEG and EMG signals.

To mitigate the above-mentioned issues and enhance the perception of temporal dynamics in EEG data and EMG data, we introduce EEG-EMG FAConformer, a novel convolutional Transformer. We propose fusing EMG features with EEG features for better recognition, an area that has received limited attention. Our model is divided into two branches. For the EMG section, we just adopted resblocks to encode information. As for the EEG section, the workflow consists of a EEG-Denoise module and an EEG branch. We would like to introduce the EEG branch completely. First, the frequency band attention will arrange an adaptive weight to different frequency which will eliminate a lot of irrelevant information. Then the Multi Scale Feature Fusion module is comprised of 1D kernels with different sizes. In that way, we can extract different hidden temporal patterns. Next the independent channel-specific convolution module is used to extract features in independent channel, which shows superior performance over normal convolution layers. Lastly, it will go through Squeeze-and-Excitation block(SEBlock) to pay more attention to related channels. Then we will fuse the EEG features and EMG features. We concat these features first and apply a fuse module to fuse these features. It will capture both short term and long term correlations hidden in signals, which is quite critical in motor pattern recognition.

Our contributions can be summarized as follows:

1. We propose a lightweight network named EEG-EMG FAConformer which is comprised of convolutions and different attention modules for motor pattern recognition decoding using EEG and EMG signals. We especially introduce a Frequency Band Module to effectively encode EEG information and elimintate noises and irrelevant information.

2. We develop some useful modules like Multi-Scale Fusion Module, Independent Channel-Specific Convolution Module (ICSCM), and Fuse Module which can effectively eliminate irrelevant information in EEG and EMG signals and fully exploit hidden patterns. Each of them can play an important part in the future work about the fusion of EEG and EMG in motor pattern recognition and greatly improve motor pattern recognition performance.

3. Through extensive experiments and ablation study on Jeong2020 dataset, the efficacy and robustness of EEG-EMG FAConformer is proved. Our results show its superiority compared with other state-of-the-art (SOTA) methods. 

\section{Related Works}
The motor pattern recognition task has a long history. It used to be mainly in machine learning. However, with the boom of deep learning, more and more methods have been proposed to enhance the decoding accuracy of motor pattern recognition.

Methods using machine learning adopt a CSP+SVM structure. Barachant et al. (2011) introduced a BCI framework using motor imagery (MI) with a focus on Riemannian geometry to classify EEG signals directly through their spatial covariance matrices. Common Spatial Pattern (CSP) \cite{blankertz2007optimizing} analyzes the spatial distribution of multi-channel EEG signals to differentiate brain activities across various tasks. Then the method utilizes svm to make a classification.

To enhance CSP's effectiveness, Optimal Spatial-Temporal Patterns (OSTP) \cite{ang2012mutual} employs mutual information for feature selection to optimize both frequency band and time segment selection for CSP filtering. Additionally, Ozdenizci et al. \cite{ozdenizci2017electroencephalographic} proposed an information-theoretic learning approach to improve neural feature interpretation and address conventional feature selection limitations.

Another advanced method, Filter Bank CSP (FBCSP) \cite{ang2012filter}, segments EEG data into multiple frequency bands and applies CSP to each. This method uses a feature selection algorithm that automatically identifies features tailored to individual subjects, significantly boosting classification accuracy. FBCSP is highly regarded and frequently used in comparative analyses within the field.

As we have mentioned above, multiple deep learning methods have been proposed to demonstrate promising results in the motor imagery task. EEGNet\cite{lawhern2018eegnet}  is the first compact deep learning model which is used in this sphere. This model is comprised of some easy modules which guarantee it is lightweight. It begins with some convolutions along the temporal dimension and then uses some depthwise convolutions along spatial fields and ends with depthwise seperable convolutions to extract features. Then some methods like FBCNet\cite{mane2021fbcnet} appears, which manually extracts temporal features by calculating variance. Since motor imagery appears mainly in some of the regions of our brains, the attention mechanism shows great performance in this task. EEG Conformer\cite{song2022eeg} which use self-attention to learn global features dominated this task. And some other methods have been proposed continuously to solve this problem. LightConvNet\cite{ma2023temporal} and some other methods have been proposed to reduce the number of parameters and increase precision. 

However, few researchers cast attention on multi-modal fusion into the motor pattern recogniton. Some researchers in the biomedical engineering field and medical field like Grosprêtre\cite{grospretre2018neural} and Decety\cite{decety1996neurophysiological} have found that when human beings move their upper limbs or do some motor imagery, related neuron groups will be activated. Inspired by these works, we analyze and fuse this information to make a more accurate estimation of one's intention for upper limb moves. 
\section{Proposed Methods}
\subsection{Overview for EEG-EMG FAConformer}
\begin{figure*}[htbp]
\begin{center}
\includegraphics[width=1\linewidth]{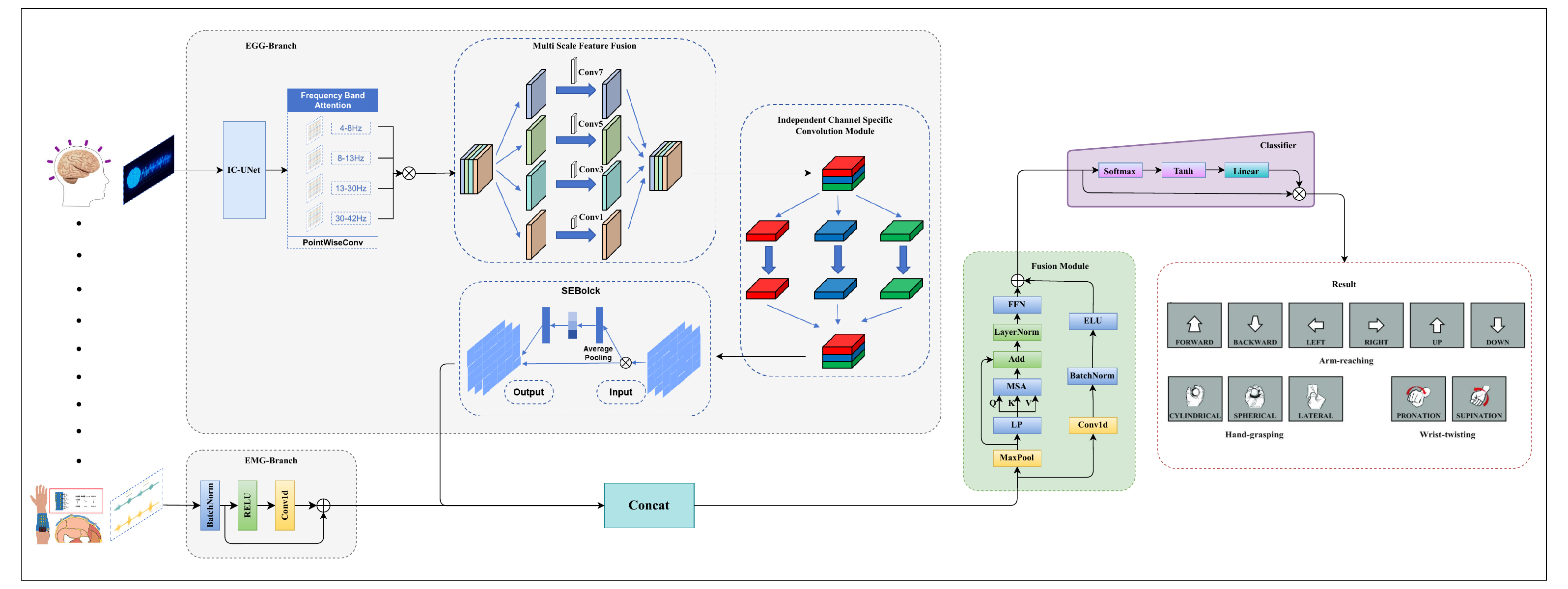}
\end{center}
\caption{\textbf{Overall framework of our network} }
\label{fig:framework}
\end{figure*}
As shown in Fig. \ref{fig:framework}, we introduce the overall structure of the model, which is divided into two branches: the EMG branch and the EEG branch. We extract information from both sources and make an effective fusion to make a more accurate motor pattern decoding.
\subsection{Details of EEG-Branch}
\subsubsection{EEG Denoise Net}
Our design for EEG-Denoise is relatively straightforward. We have incorporated the IC-U-Net model proposed by Chuang et al.\cite{chuang2022ic}, which has notably enhanced the signal-to-noise ratio of the dataset and improved classification efficiency. Despite its simplicity, this module is lightweight and demonstrates measurable improvements.
\subsubsection{Details of EEG Branch}
As shown in Fig. \ref{fig:framework}, we can get an overview of the EEG-Branch structure. Motivated by Qin\cite{qin2024m} and Ang\cite{ang2008filter}, we designed a frequency band attention module which derives valuable information from relevant frequency bands.

Initially, we adopted  multiple band-pass filters to the raw EEG data utilizing the Chebyshev Type II filter. 

Next, we used point-wise convolution to integrate multi-band information. This process allows the network to leverage complementary information from each frequency band. At the same time, an adaptive weight is assigned to each frequency band to reduce noise in redundant frequency bands and enhance information in other frequency bands.

By integrating the frequency band information, we obtain the final output. For the single-trial input data $X\in\mathbb{R}^{C\times T}$, let C represent the number of channels and T represent the time points. We applied multiple Chebyshev Type II filters to the raw data to achieve frequency splitting, resulting in data represented as $X_{MB}(n)=X*h(n)\in\mathbb{R}^{N_b\times C\times T} $ , where $N_b$ represents the number of frequency bands.

For each frequency band, we first applied a self-attention mechanism across channels. Then, we used point-wise convolution to merge the multiband information. This process enables the network to utilize the complementary information from each frequency band effectively. Simultaneously, an adaptive weight W(n) is assigned to each frequency band to attenuate noise from redundant bands and improve useful information from other bands.

After merging the frequency band information, we obtain the fused frequency band data. This approach effectively extracts valuable information from relevant frequency bands while suppressing noise and irrelevant information from other bands, leading to a refined output $X_{FS}$. And W(n) represents the weight of respective frequency.
\begin{equation}
    \begin{aligned}
     X_{FS} = \sum_{t=1}^{N_b} X_{MB}(n,i,j)*w(n)\in\mathbb{R}^{C \times T}
    \end{aligned}
\end{equation}
\

Next, we applied a set of 1D residual convolutional layers with varying kernel sizes to the extracted features for temporal feature extraction. Through experimental evaluation, we determined that the performance of 2D convolutions was inferior to that of 1D convolutions. Additionally, 1D convolutions have a relatively lower parameter count. Consequently, we utilized 1D convolutions for this purpose. By performing frequency segmentation and applying self-attention mechanisms across different channels, the primary objective is to extract features from the temporal sequence. We adopted a convolutional kernel size of \(X_s\).

Drawing inspiration from multi-scale fusion techniques, we propose a multi-scale convolutional module to learn patterns with different temporal lengths and fuse features effectively. This approach enhances recognition accuracy by employing convolutional layers of four different sizes and subsequently fusing the features to recognize patterns across various scales. Let the kernel sizes be denoted as
$S_1$,$S_2$,$S_3$ and $S_4$. The formula for extracting the features on this temporal sequence is:
\begin{equation}
    X_{LS_k}(i) = \sum_{m=0}^{S_k-1} X_{FS}(i + m) \cdot w_k(m) + b_k, \quad k=1,2,3,4
\end{equation}

\begin{equation}
    X_{\text{fuse}} = \text{concat}(X_{LS_1}, X_{LS_2}, X_{LS_3}, X_{LS_4})
\end{equation}

\begin{equation}
    X_{\text{fuse}} = \text{Conv1D}(X_{\text{fuse}}, \text{kernel\_size}=1, \text{filters}=128)
\end{equation}

These  convolutional kernels enhances our network's ability to recognize spatiotemporal information. After the fusion we adopted an independent channel-specific convolution module to lower the dimension and reduce redundant information. 

Then We utilized SEBlock\cite{hu2018squeeze} to assign different weights to different EEG channels, improving the accuracy of our classification. The operation here is also straightforward.

First, we pool the features along the temporal dimension, then learn the relationships between different channels through linear layers. This operation is part of a very lightweight gating mechanism, formulated as:

\begin{equation}
    \begin{aligned}
      Z_c = F_{sq}(X_{fuse}) = \frac{1}{T} \sum_{t=1}^{T} X_{fuse}(t)\\
      X_c = F_{ez}(Z_c, w) = \sigma(w_2 \sigma(w_1 Z_c))
    \end{aligned}
\end{equation}
This simple operation helps maintain the lightweight nature of the model.
\subsubsection{Details of EMG-Branch}
We adopt 1D convolutional residual layers to extract the EMG signals. While EMG signals can provide valuable information, their role in motor imagery tasks is primarily auxiliary. Therefore, the EMG branch is relatively smaller in comparison to the EEG branch.
\subsection{Details of Fuse Module}
In this section, we use a relatively simple method of concatenating (concat) and multihead attention mechanisms to perform feature fusion of EEG and EMG signals. We can express the signal fusion module using the following formula: 
\begin{equation}
    \begin{aligned}
      X_{fuse} = \text{MultiheadAttention}(\text{Concat}(X_{emg}, X_{eeg}))
    \end{aligned}
\end{equation}

Here, our multi-head attention mechanism reassigns weights to different EEG and EMG channels, improving the effectiveness of our classification tasks.
\section{experiments and results}
\subsection{Dataset}
The experimental data for this study was obtained from the publicly accessible Jeong2020 database \cite{jeong2020multimodal}.

The Jeong2020 dataset was recorded by a team of cognitive science professors from Korea University, including Schalk and McFarland. This dataset includes data from each participant's motor execution tasks and motor imagery tasks.

These data were provided by 25 participants, each contributing 3300 trials throughout the experiment. Each experiment included motor imagery, actual movement, and rest intervals. During the experiment, EEG signals were collected through 60 electrode channels with a sampling frequency of 2500 Hz, in addition to 6-channel EMG signals.


\subsection{Implementation Details and Evaluation Metrics}
We used multiple sets of hyperparameters in our experiments to achieve the best experimental results. We tested and recorded the number of epochs, the learning rate, the selection of the optimizer, and the batch size. Here, we use the five-fold cross-validation method for comparison, as it is commonly used in related papers. During training, we use one-fifth of the data as the test set and four-fifths as the training set, then perform five-fold calculations and average the accuracy results.

The evaluation metrics for the experiment include the most commonly used metrics for multiclassification tasks: accuracy and the kappa coefficient. According to the classification results \cite{alwasiti2020motor}, let \( T \) (true) be the number of samples where the predicted results match the actual results and \( F \) (false) be the number of samples where the predicted results differ from the actual results. The accuracy and kappa coefficient are defined as follows:

\[ \text{acc} = \frac{T}{T + F} \]

\[ \kappa = \frac{p_0 - p_e}{1 - p_e} \]

In the kappa coefficient formula, \( p_0 \) is the accuracy, used to evaluate the balance of classification. The calculation method for \( p_e \) is shown in the formula below: \( a \) represents the actual number of each class, and \( b \) represents the number of predictions for each class.

\[ p_e = \frac{a_1 \times b_1 + a_2 \times b_2 + \ldots + a_n \times b_n}{n \times n} \]

And the best hyperparameters is as follows: 
\begin{table}[htbp]
\centering
\caption{Optimal Hyperparameters}
\begin{tabular}{|c|c|}
\hline
{\textbf{Hyperparameters}} & \textbf{Corresponding number} \\ 
\hline
{\textbf{learning rate}} & \textbf{$1e^{-6}$} \\ 
{\textbf{epoch}} & {500} \\ 
{\textbf{optimizer}} & {Adam} \\ 
{\textbf{batch size}} & {100} \\ 
{\textbf{Loss}} & {Cross Entropy with R-DropOut Loss}\\
\hline
\end{tabular}
\end{table}
\begin{table*}[h]
    \centering
    \caption{Accuracy and Kappa}
    \scriptsize 
    \begin{tabular}{ccccccc}
        \toprule
        & \textbf{Ours} & \textbf{EEG Conformer}\cite{song2022eeg} & \textbf{CNN-LSTM} \cite{li2022motor} & \textbf{BaseCNN}\cite{lawhern2018eegnet} & \textbf{LMDANet}\cite{miao2023lmda} & \textbf{EEG-TCNet}\cite{ingolfsson2020eeg} \\ \midrule
        \multicolumn{7}{c}{\textbf{Accuracy}} \\ \midrule
        Multigrasp-mi & \textbf{62.5} & 56 & 43.45 & 38 & 61.5 & 59.427 \\
        Multigrasp-realmove & \textbf{94.7} & 92.6 & 91.7 & 87 & 93.5 & - \\
        Reaching-realmove & \textbf{90} & 89.1 & 88 & 85 & 89.5 & - \\
        Twist-realmove & \textbf{98.1} & 97.8 & 97.1 & 96.5 & 98 & - \\
        Twist-mi & \textbf{61.23} & 53.2 & 51.707 & 40.2 & 54.867 & 53.333 \\ \midrule
        \multicolumn{7}{c}{\textbf{Kappa}} \\ \midrule
        Multigrasp-mi & \textbf{0.437} & 0.364 & 0.150 & 0.123 & 0.414 & 0.391 \\
        Twist-mi & \textbf{0.225} & 0.123 & 0.01 & -0.2 & 0.067 & -0.007 \\ \bottomrule
    \end{tabular}
    \label{fig:data}
\end{table*}
\subsection{Experimental Results and Ablation Studies}
\subsubsection{Experimental Results}
\begin{figure*}[h]
\begin{center}
\includegraphics[width=1.0\linewidth]{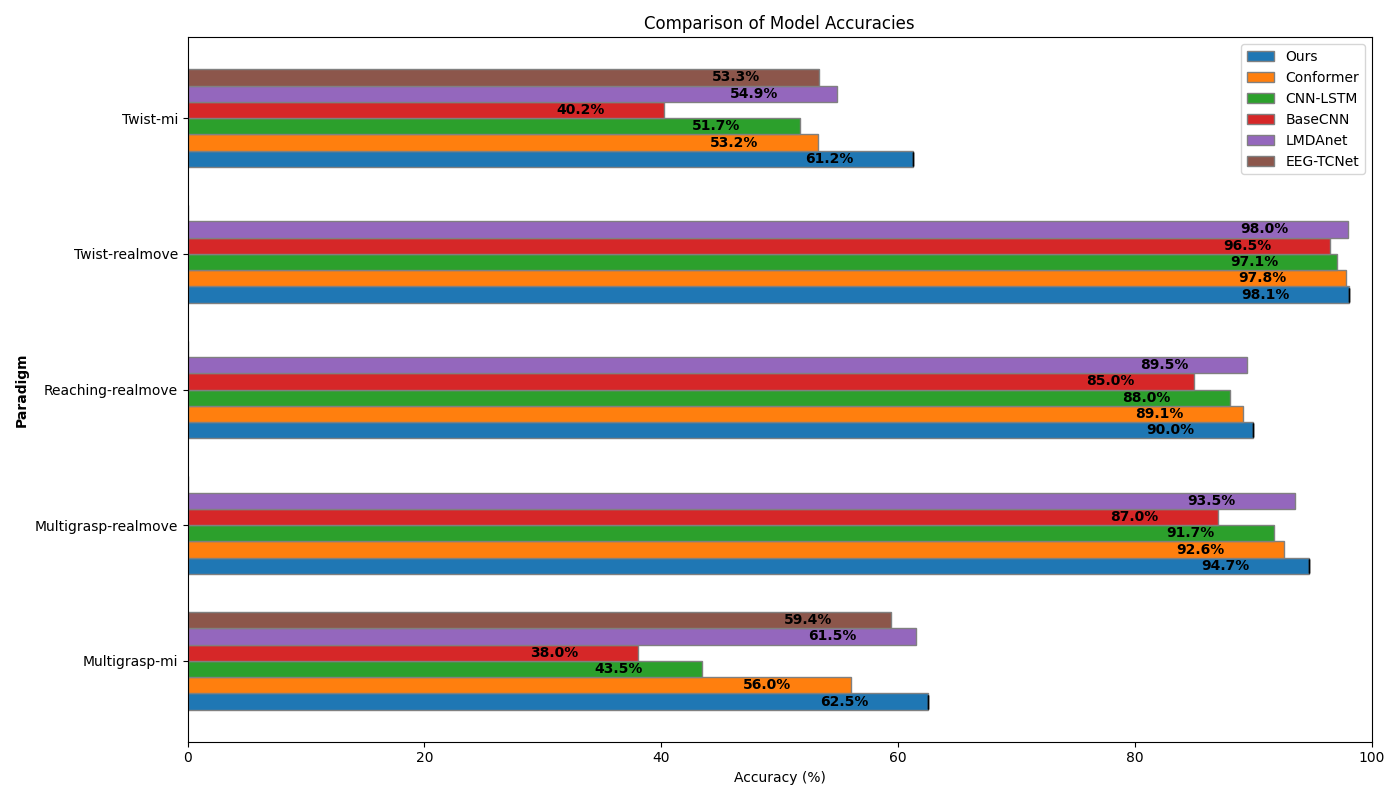}
\end{center}
\caption{\textbf{Accuracy Comparison Of Models} }
\label{fig:Accuracy}

\end{figure*}

Fig. \ref{fig:Accuracy} and Table \ref{fig:data} presents the classification accuracy results of various models across different tasks.

In the three-class motor execution task, our model achieved an accuracy of 94.7\%, significantly higher than other models. The accuracies of the other models were as follows: EEG Conformer at 92.6\%, CNN-LSTM at 91.7\%, LMDAnet at 93.5\%, and BaseCNN at 87.0\%.

In the six-class motor execution task, our method (EEG-EMG FAConformer) achieved the highest accuracy at 90.0\%. The accuracies of EEG Conformer and CNN-LSTM were 89.1\% and 88.0\%, respectively. LMDAnet achieved an accuracy of 89.5\%, and BaseCNN achieved 85.0\%.

For the binary classification task of motor execution, our method (EEG-EMG FAConformer) achieved an outstanding accuracy of 98.1\%. The accuracies of EEG Conformer and CNN-LSTM were 97.8\% and 97.1\%, respectively. LMDAnet achieved an accuracy of 98.0\%, close to our model. BaseCNN achieved 96.5\%.

In the motor imagery dataset, as shown in Fig. \ref{fig:Accuracy}, the proposed EEG-EMG FAConformer significantly outperformed in the three-class motor imagery task. EEG-EMG FAConformer achieved an accuracy of approximately 62.5\%, while EEG Conformer achieved 56\%, CNN-LSTM achieved 43.45\%, BaseCNN achieved 38\%, LMDAnet achieved 61.5\% and EEG-TCNet achieved 59.427\%.

In the binary classification task of motor imagery, the proposed EEG-EMG FAConformer achieved an accuracy of 61.23\%, while other methods lagged behind significantly. The accuracies of EEG Conformer, CNN-LSTM,EEG-TCNet and BaseCNN were only 53.2\%, 51.707\%, 53.333\% and 40.2\%, respectively. LMDAnet achieved an accuracy of 54.867\%, leading other methods but still far behind our model.

The lower classification accuracy in the motor imagery dataset is primarily due to the weak EMG signals during the data collection process, as the muscles were only slightly activated. In motor execution, EMG signals decode movements more effectively, making decoding easier. Therefore, we plan to create a motor imagery dataset with slightly increased force to address the issue of relatively weak EMG signals.

These comparisons demonstrate that our method (EEG-EMG FAConformer) excels in multiple tasks, with leading accuracies, highlighting its strong feature extraction and classification performance. LMDAnet also performed well in some tasks, but overall, it still lagged behind our method. These results further validate the effectiveness and robustness of our method in handling complex tasks.

\subsubsection{Visualization}
\begin{figure}[h]
    \centering
    \begin{subfigure}[b]{0.48\linewidth}
        \centering
        \includegraphics[width=\linewidth]{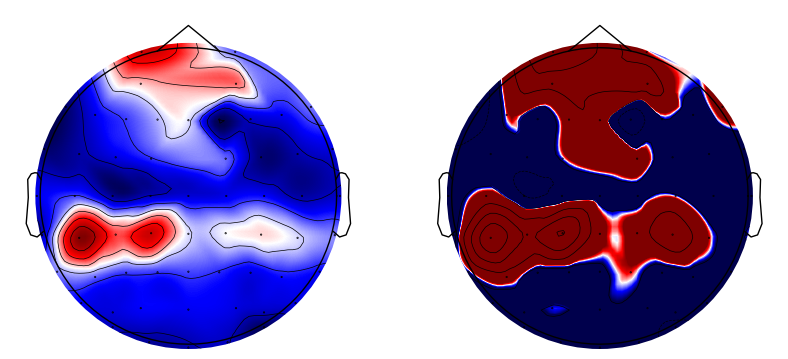} 
        \caption{Subject2}
        \label{fig:head2}
    \end{subfigure}
    \hfill
    \begin{subfigure}[b]{0.48\linewidth}
        \centering
        \includegraphics[width=\linewidth]{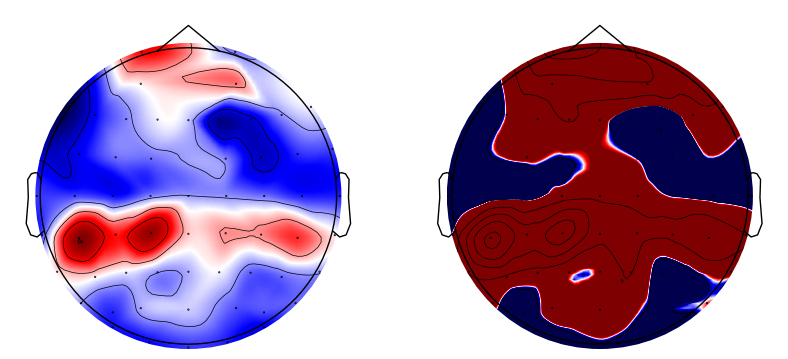} 
        \caption{Subject3}
        \label{fig:head3}
    \end{subfigure}
    \hfill
    \begin{subfigure}[b]{0.48\linewidth}
        \centering
        \includegraphics[width=\linewidth]{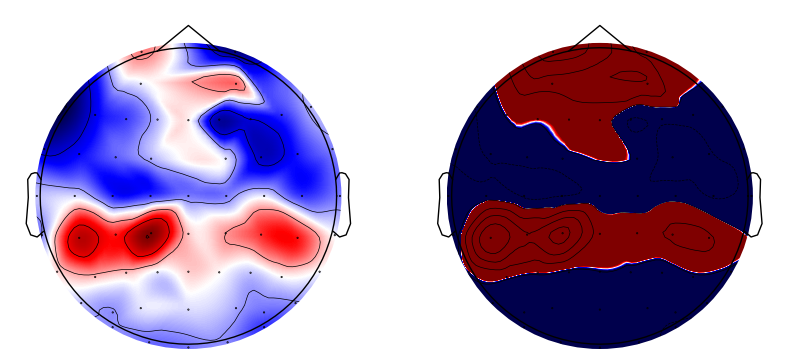} 
        \caption{Subject4}
        \label{fig:head4}
    \end{subfigure}
    \hfill
    \begin{subfigure}[b]{0.48\linewidth}
        \centering
        \includegraphics[width=\linewidth]{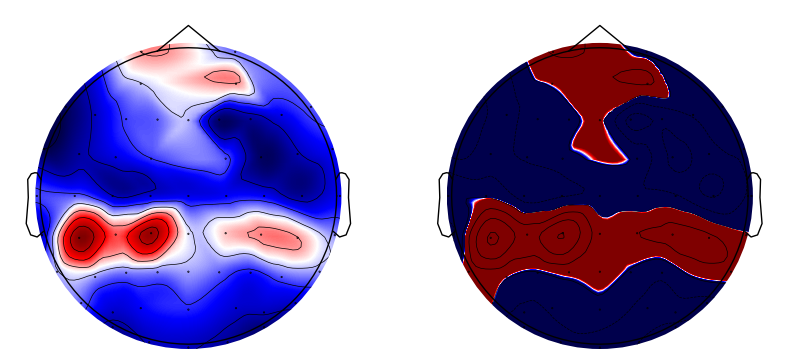} 
        \caption{Subject5}
        \label{fig:head5}
    \end{subfigure}
    \hfill
    \begin{subfigure}[b]{0.48\linewidth}
        \centering
        \includegraphics[width=\linewidth]{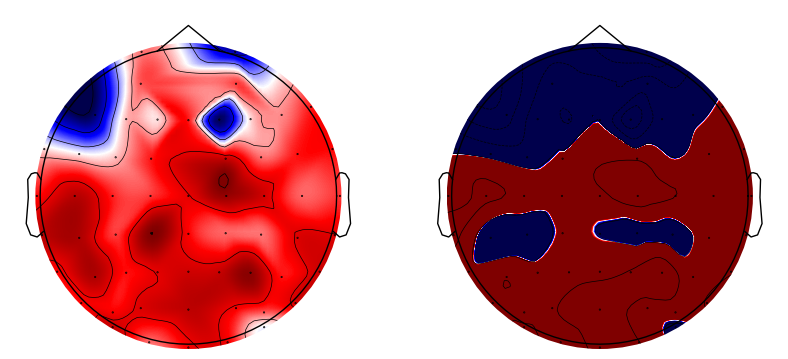} 
        \caption{Subject6}
        \label{fig:head6}
    \end{subfigure}
    \begin{subfigure}[b]{0.48\linewidth}
        \centering
        \includegraphics[width=\linewidth]{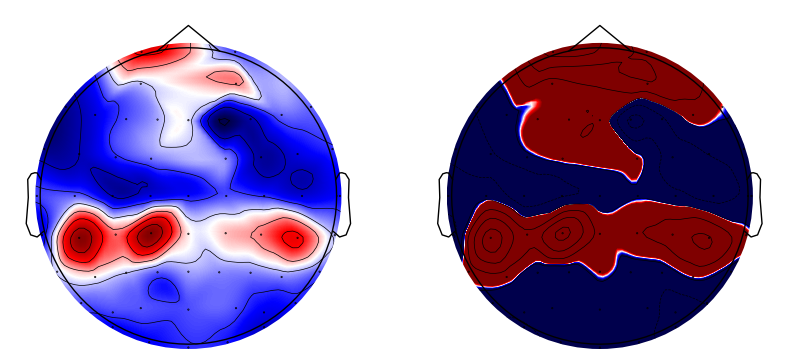} 
        \caption{Subject7}
        \label{fig:head7}
    \end{subfigure}
    \caption{\textbf{Visualization of heads for multigrasp-mi: RAW and Features EEG-EMG FAConfomer extracts}}
    \label{fig:head_comparison}
\end{figure}

Fig. \ref{fig:head_comparison} shows the raw brain region images (left side) and the brain region visualizations processed by our frequency-band attention mechanism (right side) for different subjects (Subject 2 to Subject 7). These images demonstrate the significant effect of our method in enhancing feature extraction and classification tasks.

After processing with our method, it is evident that the features of the relevant activated regions are significantly enhanced. Specifically, in the raw data, the distribution of the activated regions is not very distinct. However, after applying the frequency-band attention mechanism, the signal intensity in these regions is also enhanced.

These improvements are highly beneficial for classification tasks. Enhanced features help the model better identify and distinguish different categories, thereby improving classification accuracy. The comparison of data for each subject in Fig. \ref{fig:head_comparison} clearly shows the effectiveness of our frequency-band attention mechanism in extracting key features.

Further ablation study also verified the role of the frequency band attention mechanism. In the ablation experiments, when the frequency band attention mechanism was removed, the model's performance significantly declined, demonstrating the critical role of the attention mechanism in feature extraction and enhancement. These experimental results not only showcase the superiority of our method but also provide strong support for its effectiveness in practical applications.

Fig. \ref{fig:head_comparison} shows significant differences in motor imagery EEG patterns among subjects, leading to lower test accuracies for some individuals. For example, Subject 6 exhibits distinct brain region patterns compared to others. This phenomenon can be attributed to several factors: first, variations in adherence to experimental guidelines and environmental factors might have compromised EEG signal quality. Equipment accuracy and stability also play a role. Second, EEG signals are highly individualized, influenced by physiological and psychological states, and motor imagery abilities. Some subjects may have weaker or atypical motor imagery patterns, impacting the classifier's performance. These individual differences pose a significant challenge in motor imagery EEG signal classification, as varying motor imagery abilities and styles among subjects result in lower classification accuracy. To address this, future research should focus on improving data collection to enhance signal quality and consistency, developing robust algorithms for feature extraction and classification to better handle individual differences, and implementing personalized modeling approaches to customize training based on unique EEG characteristics, thereby improving classification accuracy.
\begin{figure}[htbp]
    \centering
    \begin{subfigure}[b]{0.48\linewidth}
        \centering
        \includegraphics[width=\linewidth]{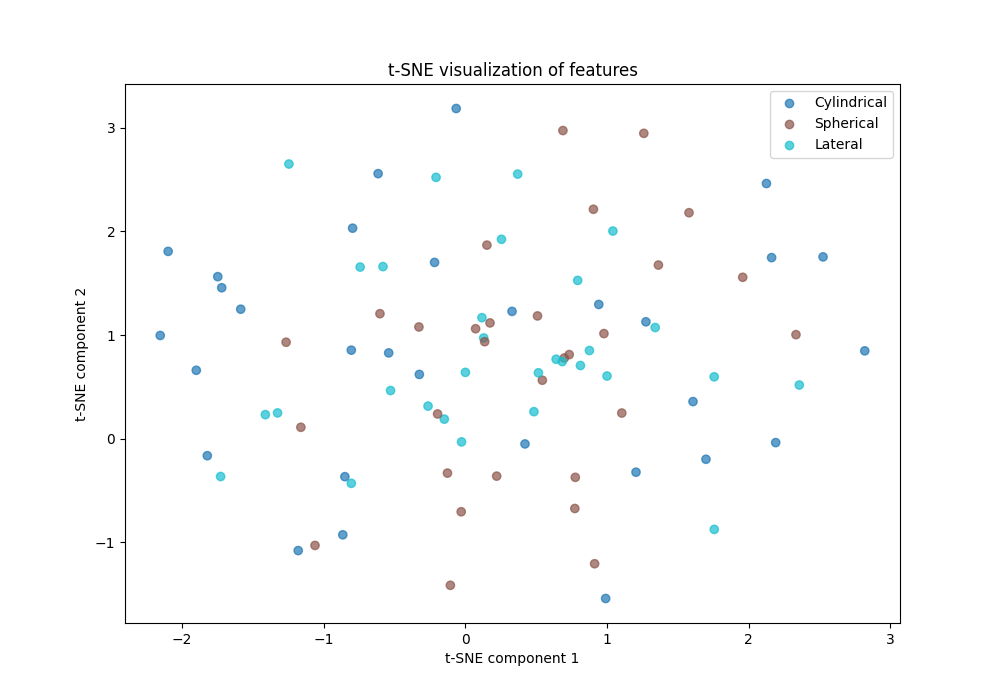} 
        \caption{Original Features}
        \label{fig:original}
    \end{subfigure}
    \hfill
    \begin{subfigure}[b]{0.48\linewidth}
        \centering
        \includegraphics[width=\linewidth]{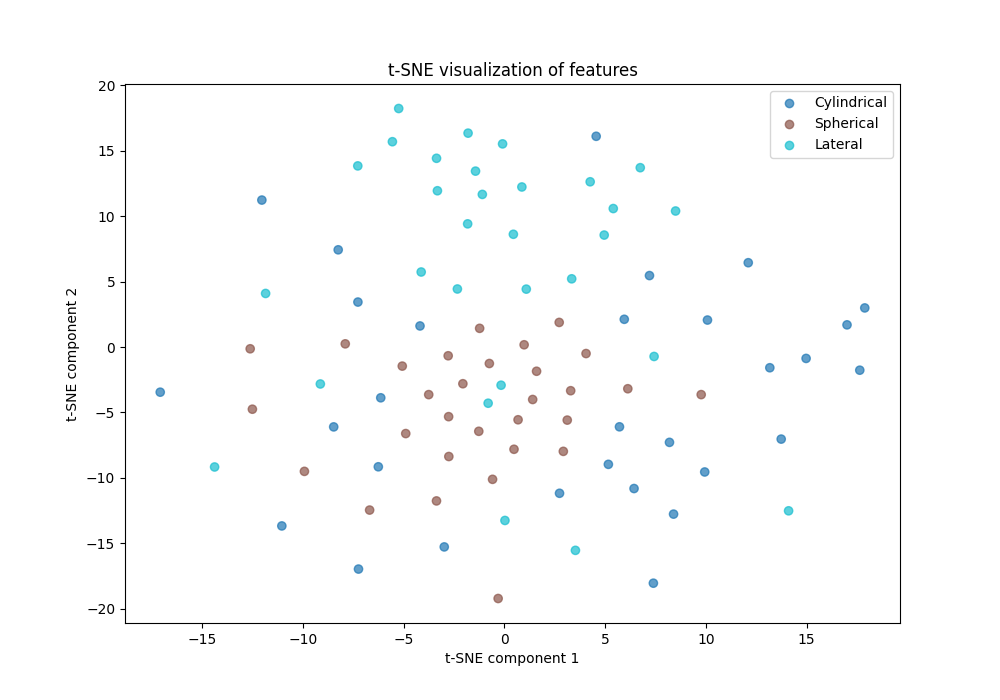} 
        \caption{LMDAnet}
        \label{fig:LMDAnet}
    \end{subfigure}
    \hfill
    \begin{subfigure}[b]{0.48\linewidth}
        \centering
        \includegraphics[width=\linewidth]{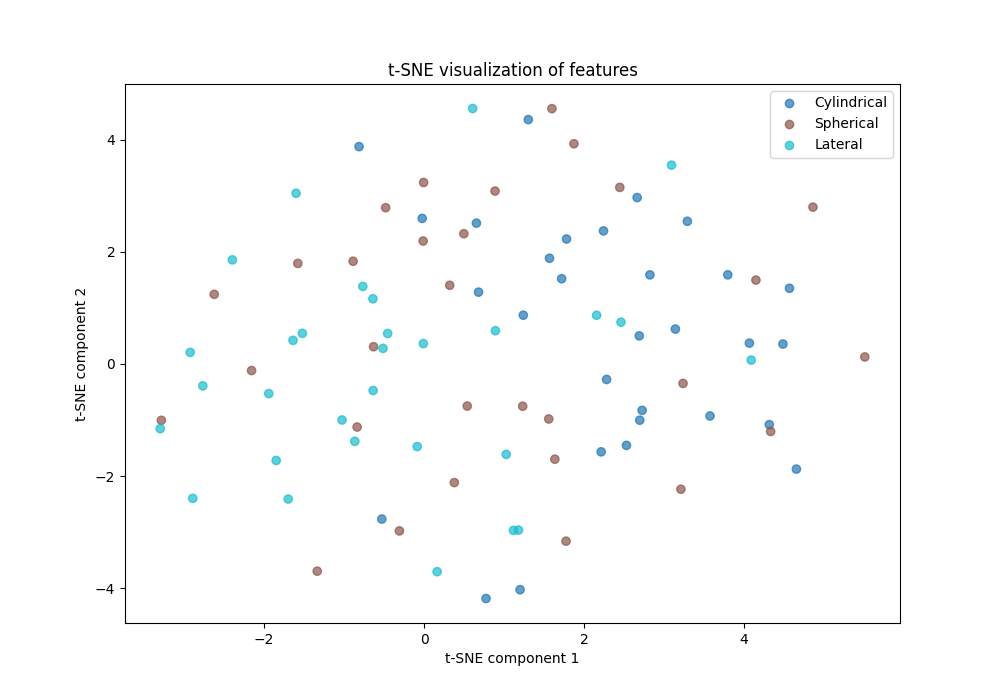} 
        \caption{LSTMCNN}
        \label{fig:lstmcnn}
    \end{subfigure}
    \hfill
    \begin{subfigure}[b]{0.48\linewidth}
        \centering
        \includegraphics[width=\linewidth]{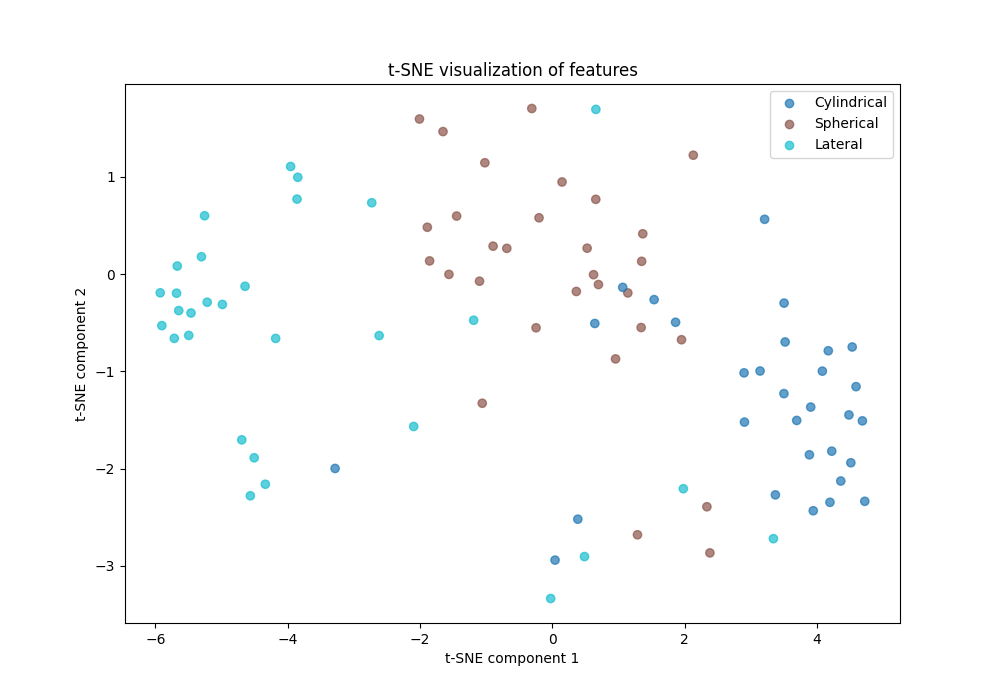} 
        \caption{Ours}
        \label{fig:ours}
    \end{subfigure}

    \caption{t-SNE Dimensionality Reduction Comparison of Multigrasp-mi}
    \label{fig:t-sne1}
\end{figure}
\begin{figure}[htbp]
    \centering
    \begin{subfigure}[b]{0.48\linewidth}
        \centering
        \includegraphics[width=\linewidth]{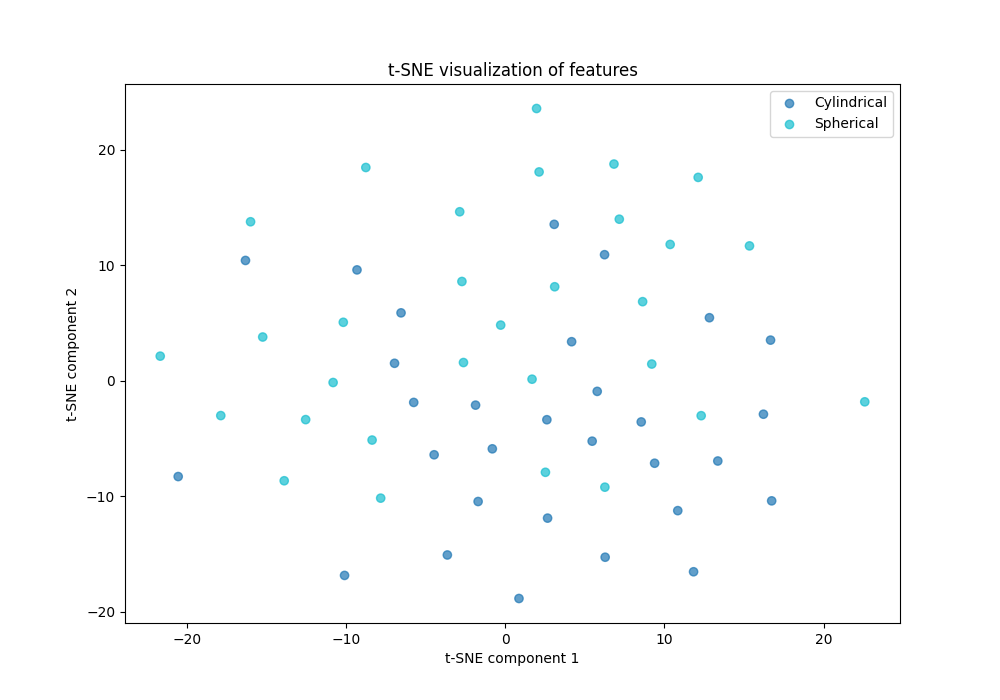} 
        \caption{Original Features}
        \label{fig:original}
    \end{subfigure}
    \hfill
    \begin{subfigure}[b]{0.48\linewidth}
        \centering
        \includegraphics[width=\linewidth]{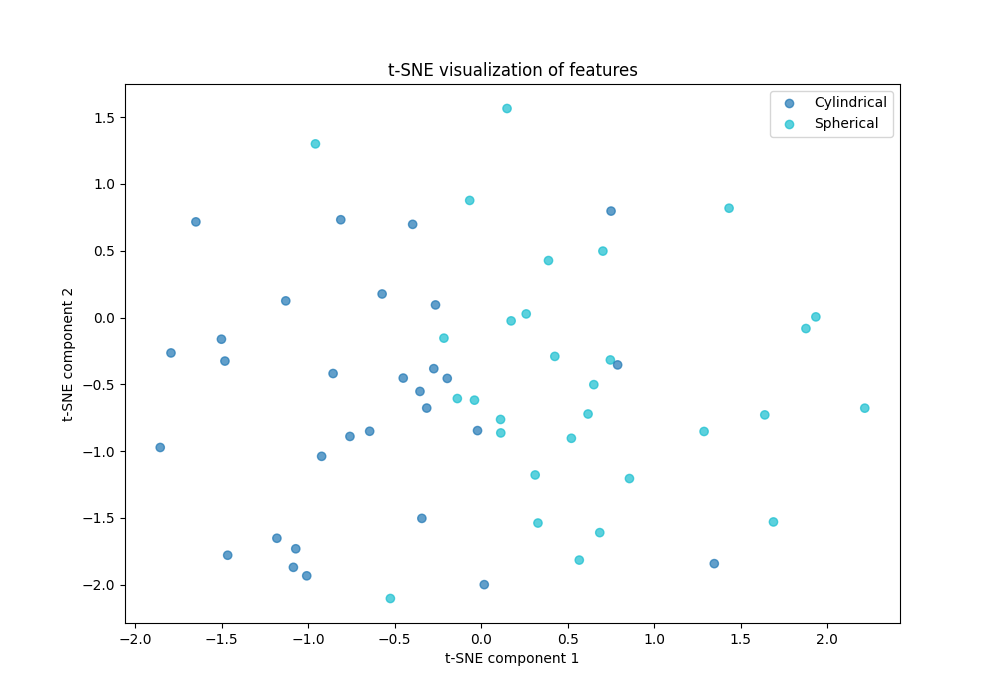} 
        \caption{LMDAnet}
        \label{fig:LMDAnet}
    \end{subfigure}
    \hfill
    \begin{subfigure}[b]{0.48\linewidth}
        \centering
        \includegraphics[width=\linewidth]{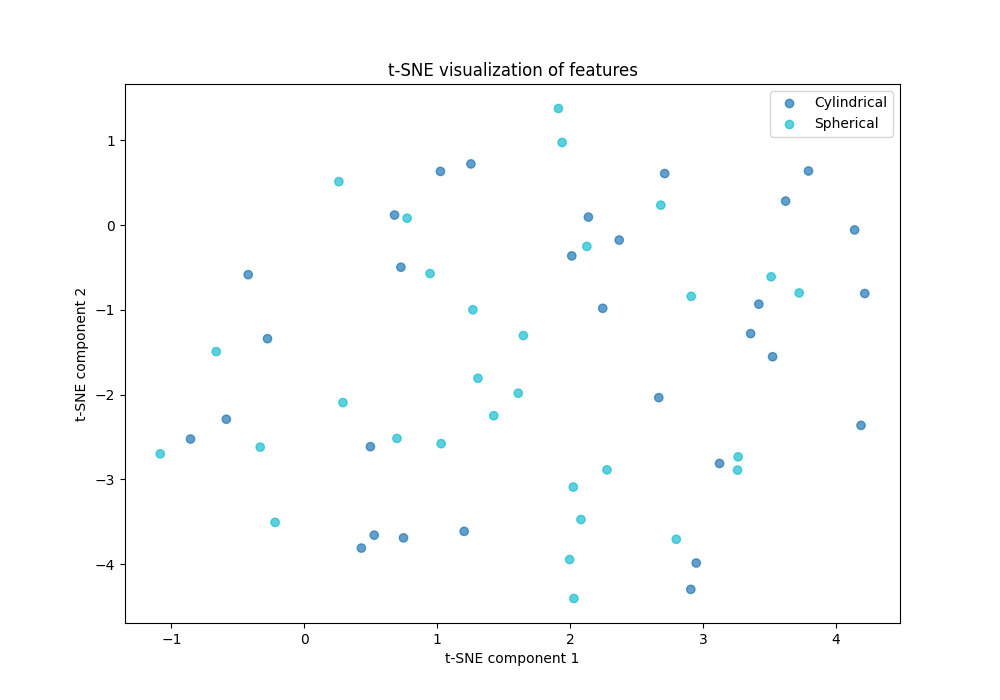} 
        \caption{LSTMCNN}
        \label{fig:lstmcnn}
    \end{subfigure}
    \hfill
    \begin{subfigure}[b]{0.48\linewidth}
        \centering
        \includegraphics[width=\linewidth]{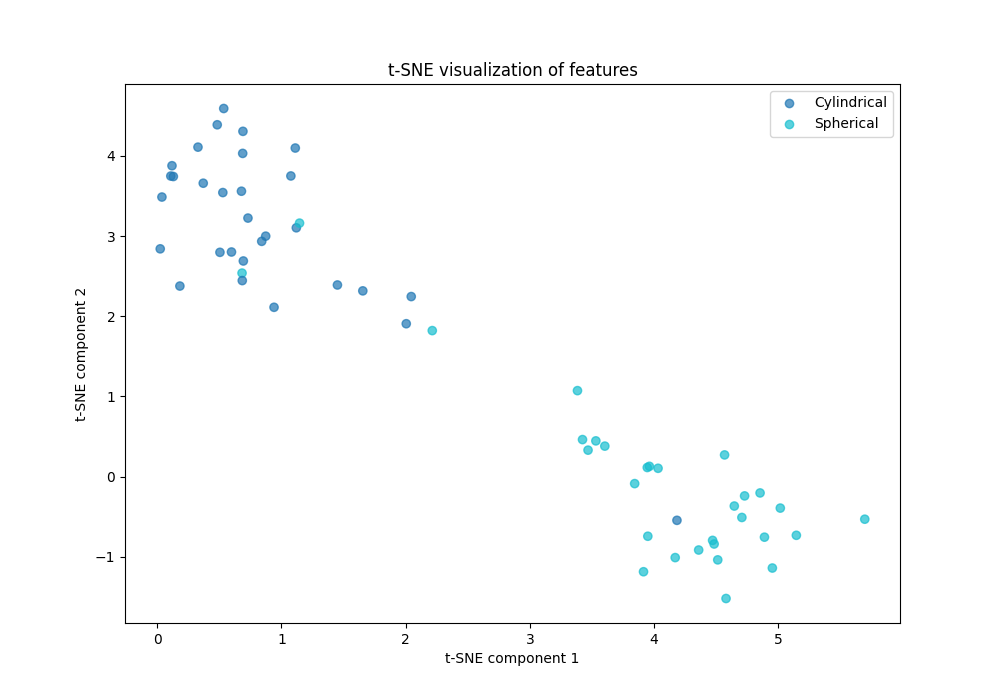} 
        \caption{Ours}
        \label{fig:ours}
    \end{subfigure}
    \caption{t-SNE Dimensionality Reduction Comparison of Twist-mi}
    \label{fig:t-sne2}
\end{figure}

t-Distributed Stochastic Neighbor Embedding (t-SNE)\cite{van2008visualizing} is a commonly used statistical dimensionality reduction and feature visualization method. After training with LMDAnet, LSTMCNN, and our method, Fig. \ref{fig:t-sne1} and \ref{fig:t-sne2} show the t-SNE projections of the data into the two paradigms, 2D S01 and 2D S11, respectively. The visual representations in Fig. \ref{fig:t-sne1} and \ref{fig:t-sne2} highlight the superior feature extraction capabilities of the EEG-EMG FAConformer model, providing a reasonable explanation for its superior classification performance.

As seen in Fig. \ref{fig:t-sne1} and \ref{fig:t-sne2}, the features extracted by the EEG-EMG FAConformer model exhibit clear boundaries and more distinct clustering, indicating that this model effectively captures sparse and distinct features, enhancing the separability between different categories. Specifically, compared to other models, the EEG-EMG FAConformer can achieve better classification results with fewer training samples. These advantages are not only visually evident but also quantitatively supported by metrics such as accuracy and kappa coefficient, where the EEG-EMG FAConformer model performs better.

In contrast, the LSTMCNN and LMDAnet models show more chaotic feature extraction with unclear boundaries and poor class distinction. This indicates that the EEG-EMG FAConformer model can better utilize the critical information in EEG and EMG signals during feature extraction, thereby enhancing overall classification performance.

In conclusion, the visual results and quantitative analysis from Fig. \ref{fig:t-sne1} and \ref{fig:t-sne2} demonstrate the superiority of the EEG-EMG FAConformer model in feature extraction and classification tasks. This model not only improves the efficiency of feature extraction but also maintains high classification accuracy with fewer training samples. This is significant for practical applications, as obtaining large amounts of labeled data is often challenging in real-world scenarios. Therefore, the superior characteristics of the EEG-EMG FAConformer model provide strong support for multimodal data processing in complex scenarios.
\begin{figure}[htbp]
    \centering
    \begin{subfigure}[b]{0.48\linewidth}
        \centering
        \includegraphics[width=\linewidth]{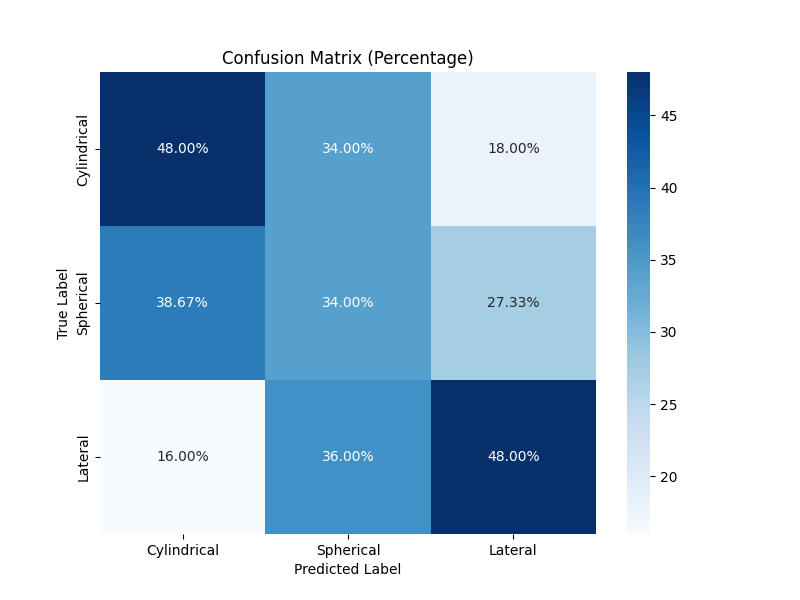} 
        \caption{CNNLSTM}
        \label{fig:Cnnlstm}
    \end{subfigure}
    \hfill
    \begin{subfigure}[b]{0.48\linewidth}
        \centering
        \includegraphics[width=\linewidth]{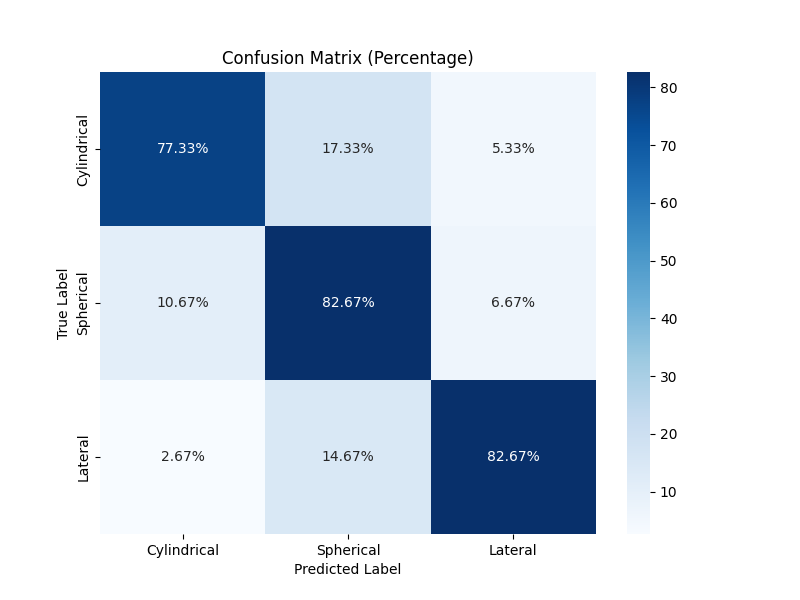} 
        \caption{LMDANet}
        \label{fig:Lmdanet}
    \end{subfigure}
    \hfill
    \begin{subfigure}[b]{0.48\linewidth}
        \centering
        \includegraphics[width=\linewidth]{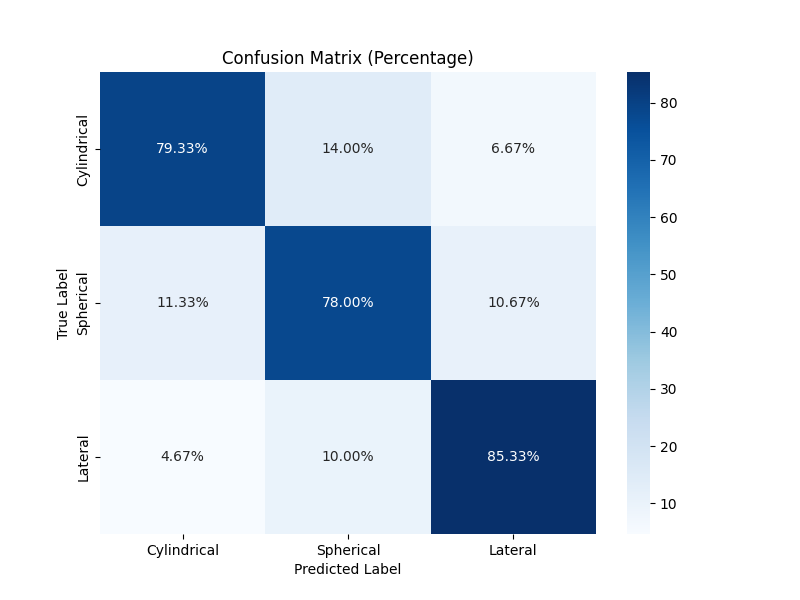} 
        \caption{Ours}
        \label{fig:Ours}
    \end{subfigure}
    \hfill
    \begin{subfigure}[b]{0.48\linewidth}
        \centering
        \includegraphics[width=\linewidth]{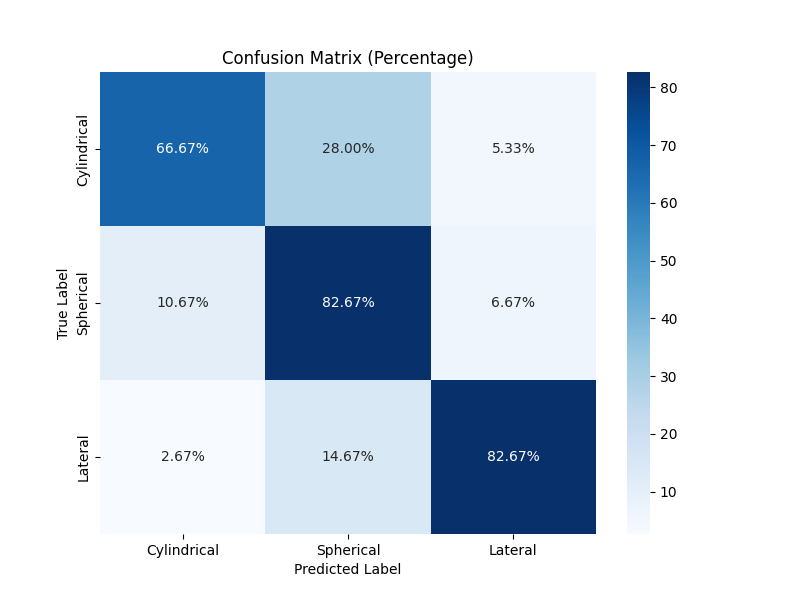} 
        \caption{EEG Conformer}
        \label{fig:matrix-conformer}
    \end{subfigure}
    \caption{Confusion matrix for three-class motor imagery classification (Subject 8)}
    \label{fig:confusion_matrix}
\end{figure}
\begin{table*}[htbp]
    \centering
    \caption{Ablation Study Based on Motor Imagery Ternary Classification}
    \begin{tabular}{|c|c|c|c|c|}
        \hline
        \multicolumn{4}{|c|}{Modules} & \multicolumn{1}{c|}{Model Accuracy (\%)} \\
        \hline
        Frequency Band Attention & Multiscale Fusion & With EMG & Independent Channel-Specific Convolution Module & \\
        \hline
         & \checkmark & \checkmark &  & 59.563 \\
         \checkmark & & \checkmark & \checkmark & 62.25 \\
         & \checkmark & \checkmark & \checkmark & 60.51 \\
         \checkmark & \checkmark & \checkmark & & 62.3 \\
         \checkmark & \checkmark & & \checkmark & 61 \\
         \checkmark & \checkmark & \checkmark & \checkmark & \textbf{62.5} \\
        \hline
    \end{tabular}
    \label{table:ablation}
\end{table*}

Fig. \ref{fig:confusion_matrix} shows the confusion matrices of four models on the three-class motor imagery task for Subject 8, including CNNLSTM (2022), LMDANet (2023), EEG Conformer (2022), and our proposed method. The confusion matrices are presented in percentage form, displaying the performance of each model in recognizing the three motor imagery classes (Cylindrical, Spherical, and Lateral).

The confusion matrix of the CNNLSTM model shows that the recognition accuracy for the Cylindrical class is 48.00\%, for the Spherical class is 34.00\%, and for the Lateral class is 48.00\%. This model has a high error rate and considerable confusion between the Cylindrical and Spherical classes.

The LMDANet model achieves a recognition accuracy of 77.33\% for the Cylindrical class, 82.67\% for the Spherical class, and 82.67\% for the Lateral class. Overall, this model performs quite balanced across the three motor imagery classes, with high recognition accuracy.

Our proposed method achieves a recognition accuracy of 79.33\% for the Cylindrical class, 78.00\% for the Spherical class, and 85.33\% for the Lateral class. This method performs balanced across the three motor imagery classes, with the highest recognition accuracy for the Lateral class.

The EEG Conformer model achieves a recognition accuracy of 66.67\% for the Cylindrical class, 82.67\% for the Spherical class, and 82.67\% for the Lateral class. This model performs excellently on the Spherical and Lateral classes but has relatively lower recognition accuracy for the Cylindrical class.

From the confusion matrices, it can be seen that our method outperforms the other three models in the three-class motor imagery tasks. The accuracy of our model is relatively stable across all classification scenarios, making it highly practical for real-world applications.
\subsection{Ablation Study}
\begin{figure}[htbp]
    \centering
    \includegraphics[width=0.5\textwidth]{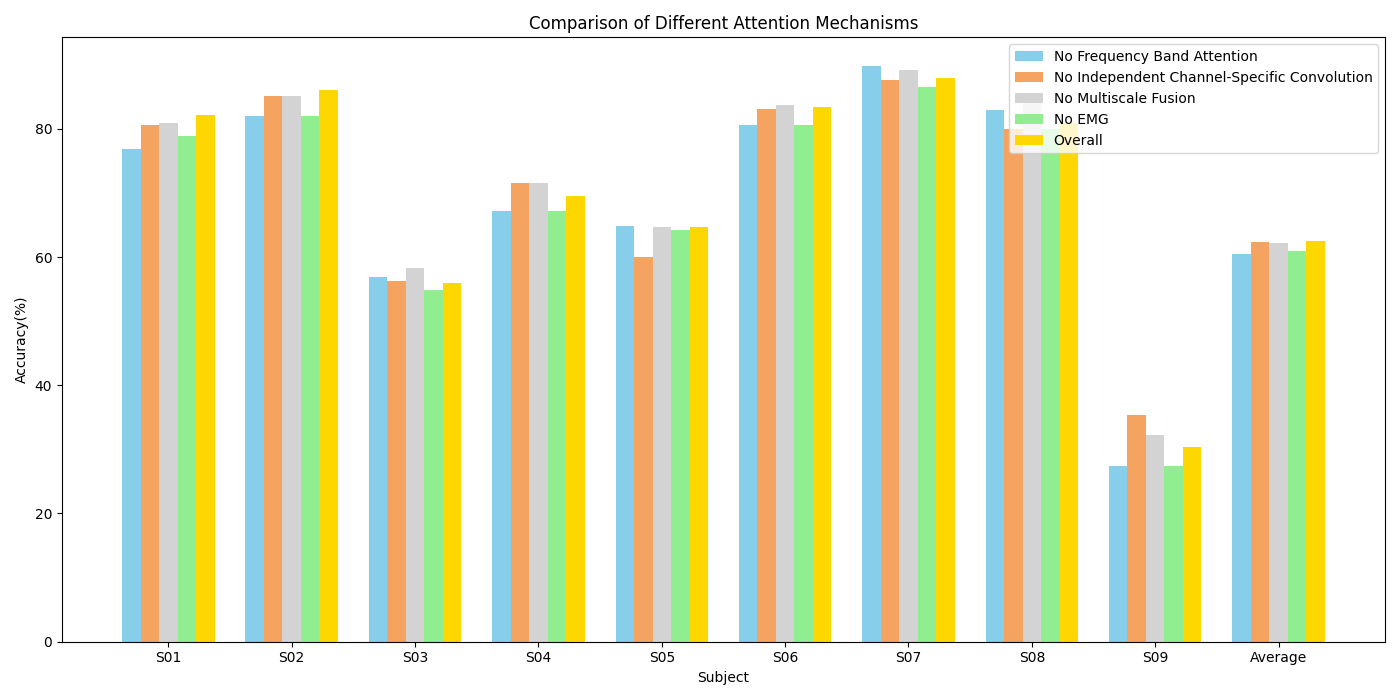}
    \caption{Comparison of Ablation Study}
    \label{fig:ablation}
\end{figure}

In this study, we conducted an ablation study on the multi-feature attention module of the EEG-EMG FAConformer to evaluate its impact on model performance. We used the Jeong2020 dataset for the experiments, specifically removing the frequency band attention, multiscale fusion and independent channel-specific convolution module. To compare the overall model with each module, we selected data from 9 participants for detailed display. The experimental results are shown in Fig. \ref{fig:ablation} and Table \ref{table:ablation}.

Specifically, after removing the modules, the accuracy of the model decreased to varying degrees. After removing the frequency band attention module, the average accuracy decreased by 2\%. After removing the multiscale fusion module, the average accuracy decreased by 0.25\%. After removing the independent channel-specific convolution module, the average accuracy decreased by 0.2\%. In almost every subject's experimental results, removing the frequency band attention module resulted in a significant loss in accuracy, such as a 6\% drop in S01. This indicates that the frequency band attention module plays an important role in improving the classification accuracy for specific tasks, particularly in motor imagery tasks.

\section{Conclusion}
In this work we have introduced a lightweight and effective model tailored for the extraction and fusion of EEG and EMG signals to recognize motor patterns. With the combination of Frequency Band Attention, Multi-Scale Fusion, Independent Channel-Specific Convolution Module (ICSCM) and Fuse Module, we effectively extract important information for recognition and get rid of redundant and irrelevant information. Our method shows astonishing robustness and superiority in Jeong2020 dataset and then verify the potential of EEG-EMG fusion in motor pattern recognition.

However, there is still a lot to be done. Firstly, there is no enough dataset with both EEG and EMG recorded in relevant tasks, which make the development of fusion of EEG and EMG signals quite difficult. Still the transfer learning in relevant area remains undeveloped. Since different people have different EEG signals, the application of transfer learning in the decoding of EEG signals is quite important.

Lastly, we are currently developing a dataset focused on mild muscle activation during motor imagery to address the problem of low EMG intensity in previous motor imagery datasets. This dataset aims to provide more diverse and realistic data, enhancing the training and performance of EEG-EMG fusion models in capturing subtle muscle activities.

\bibliographystyle{IEEEtran}
\bibliography{ref}

\begin{thebibliography}{10}
\providecommand{\url}[1]{#1}
\csname url@samestyle\endcsname
\providecommand{\newblock}{\relax}
\providecommand{\bibinfo}[2]{#2}
\providecommand{\BIBentrySTDinterwordspacing}{\spaceskip=0pt\relax}
\providecommand{\BIBentryALTinterwordstretchfactor}{4}
\providecommand{\BIBentryALTinterwordspacing}{\spaceskip=\fontdimen2\font plus
\BIBentryALTinterwordstretchfactor\fontdimen3\font minus \fontdimen4\font\relax}
\providecommand{\BIBforeignlanguage}[2]{{%
\expandafter\ifx\csname l@#1\endcsname\relax
\typeout{** WARNING: IEEEtran.bst: No hyphenation pattern has been}%
\typeout{** loaded for the language `#1'. Using the pattern for}%
\typeout{** the default language instead.}%
\else
\language=\csname l@#1\endcsname
\fi
#2}}
\providecommand{\BIBdecl}{\relax}
\BIBdecl

\bibitem{lotze2006motor}
M.~Lotze and U.~Halsband, ``Motor imagery,'' \emph{Journal of Physiology-paris}, vol.~99, no. 4-6, pp. 386--395, 2006.

\bibitem{vaughan2006wadsworth}
T.~M. Vaughan, D.~J. McFarland, G.~Schalk, W.~A. Sarnacki, D.~J. Krusienski, E.~W. Sellers, and J.~R. Wolpaw, ``The wadsworth bci research and development program: at home with bci,'' \emph{IEEE transactions on neural systems and rehabilitation engineering}, vol.~14, no.~2, pp. 229--233, 2006.

\bibitem{pfurtscheller1999event}
G.~Pfurtscheller and F.~L. Da~Silva, ``Event-related eeg/meg synchronization and desynchronization: basic principles,'' \emph{Clinical neurophysiology}, vol. 110, no.~11, pp. 1842--1857, 1999.

\bibitem{liu2020parallel}
X.~Liu, Y.~Shen, J.~Liu, J.~Yang, P.~Xiong, and F.~Lin, ``Parallel spatial--temporal self-attention cnn-based motor imagery classification for bci,'' \emph{Frontiers in neuroscience}, vol.~14, p. 587520, 2020.

\bibitem{fan2021bilinear}
C.-C. Fan, H.~Yang, Z.-G. Hou, Z.-L. Ni, S.~Chen, and Z.~Fang, ``Bilinear neural network with 3-d attention for brain decoding of motor imagery movements from the human eeg,'' \emph{Cognitive Neurodynamics}, vol.~15, pp. 181--189, 2021.

\bibitem{song2022eeg}
Y.~Song, Q.~Zheng, B.~Liu, and X.~Gao, ``Eeg conformer: Convolutional transformer for eeg decoding and visualization,'' \emph{IEEE Transactions on Neural Systems and Rehabilitation Engineering}, vol.~31, pp. 710--719, 2022.

\bibitem{she2023improved}
Q.~She, T.~Chen, F.~Fang, J.~Zhang, Y.~Gao, and Y.~Zhang, ``Improved domain adaptation network based on wasserstein distance for motor imagery eeg classification,'' \emph{IEEE Transactions on Neural Systems and Rehabilitation Engineering}, vol.~31, pp. 1137--1148, 2023.

\bibitem{li2022motor}
H.~Li, M.~Ding, R.~Zhang, and C.~Xiu, ``Motor imagery eeg classification algorithm based on cnn-lstm feature fusion network,'' \emph{Biomedical signal processing and control}, vol.~72, p. 103342, 2022.

\bibitem{lawhern2018eegnet}
V.~J. Lawhern, A.~J. Solon, N.~R. Waytowich, S.~M. Gordon, C.~P. Hung, and B.~J. Lance, ``Eegnet: a compact convolutional neural network for eeg-based brain--computer interfaces,'' \emph{Journal of neural engineering}, vol.~15, no.~5, p. 056013, 2018.

\bibitem{ingolfsson2020eeg}
T.~M. Ingolfsson, M.~Hersche, X.~Wang, N.~Kobayashi, L.~Cavigelli, and L.~Benini, ``Eeg-tcnet: An accurate temporal convolutional network for embedded motor-imagery brain--machine interfaces,'' in \emph{2020 IEEE International Conference on Systems, Man, and Cybernetics (SMC)}.\hskip 1em plus 0.5em minus 0.4em\relax IEEE, 2020, pp. 2958--2965.

\bibitem{miao2023lmda}
Z.~Miao, M.~Zhao, X.~Zhang, and D.~Ming, ``Lmda-net: A lightweight multi-dimensional attention network for general eeg-based brain-computer interfaces and interpretability,'' \emph{NeuroImage}, vol. 276, p. 120209, 2023.

\bibitem{blankertz2007optimizing}
B.~Blankertz, R.~Tomioka, S.~Lemm, M.~Kawanabe, and K.-R. Muller, ``Optimizing spatial filters for robust eeg single-trial analysis,'' \emph{IEEE Signal processing magazine}, vol.~25, no.~1, pp. 41--56, 2007.

\bibitem{ang2012mutual}
K.~K. Ang, Z.~Y. Chin, H.~Zhang, and C.~Guan, ``Mutual information-based selection of optimal spatial--temporal patterns for single-trial eeg-based bcis,'' \emph{Pattern Recognition}, vol.~45, no.~6, pp. 2137--2144, 2012.

\bibitem{ozdenizci2017electroencephalographic}
O.~{\"O}zdenizci, M.~Yal{\c{c}}{\i}n, A.~Erdo{\u{g}}an, V.~Pato{\u{g}}lu, M.~Grosse-Wentrup, and M.~{\c{C}}etin, ``Electroencephalographic identifiers of motor adaptation learning,'' \emph{Journal of neural engineering}, vol.~14, no.~4, p. 046027, 2017.

\bibitem{ang2012filter}
K.~K. Ang, Z.~Y. Chin, C.~Wang, C.~Guan, and H.~Zhang, ``Filter bank common spatial pattern algorithm on bci competition iv datasets 2a and 2b,'' \emph{Frontiers in neuroscience}, vol.~6, p. 21002, 2012.

\bibitem{mane2021fbcnet}
R.~Mane, E.~Chew, K.~Chua, K.~K. Ang, N.~Robinson, A.~P. Vinod, S.-W. Lee, and C.~Guan, ``Fbcnet: A multi-view convolutional neural network for brain-computer interface,'' \emph{arXiv preprint arXiv:2104.01233}, 2021.

\bibitem{ma2023temporal}
X.~Ma, W.~Chen, Z.~Pei, J.~Liu, B.~Huang, and J.~Chen, ``A temporal dependency learning cnn with attention mechanism for mi-eeg decoding,'' \emph{IEEE Transactions on Neural Systems and Rehabilitation Engineering}, 2023.

\bibitem{grospretre2018neural}
S.~Grospr{\^e}tre, T.~Jacquet, F.~Lebon, C.~Papaxanthis, and A.~Martin, ``Neural mechanisms of strength increase after one-week motor imagery training,'' \emph{European journal of sport science}, vol.~18, no.~2, pp. 209--218, 2018.

\bibitem{decety1996neurophysiological}
J.~Decety, ``The neurophysiological basis of motor imagery,'' \emph{Behavioural brain research}, vol.~77, no. 1-2, pp. 45--52, 1996.

\bibitem{chuang2022ic}
C.-H. Chuang, K.-Y. Chang, C.-S. Huang, and T.-P. Jung, ``Ic-u-net: a u-net-based denoising autoencoder using mixtures of independent components for automatic eeg artifact removal,'' \emph{NeuroImage}, vol. 263, p. 119586, 2022.

\bibitem{qin2024m}
Y.~Qin, B.~Yang, S.~Ke, P.~Liu, F.~Rong, and X.~Xia, ``M-fanet: Multi-feature attention convolutional neural network for motor imagery decoding,'' \emph{IEEE Transactions on Neural Systems and Rehabilitation Engineering}, 2024.

\bibitem{ang2008filter}
K.~K. Ang, Z.~Y. Chin, H.~Zhang, and C.~Guan, ``Filter bank common spatial pattern (fbcsp) in brain-computer interface,'' in \emph{2008 IEEE international joint conference on neural networks (IEEE world congress on computational intelligence)}.\hskip 1em plus 0.5em minus 0.4em\relax IEEE, 2008, pp. 2390--2397.

\bibitem{hu2018squeeze}
J.~Hu, L.~Shen, and G.~Sun, ``Squeeze-and-excitation networks,'' in \emph{Proceedings of the IEEE conference on computer vision and pattern recognition}, 2018, pp. 7132--7141.

\bibitem{jeong2020multimodal}
J.-H. Jeong, J.-H. Cho, K.-H. Shim, B.-H. Kwon, B.-H. Lee, D.-Y. Lee, D.-H. Lee, and S.-W. Lee, ``Multimodal signal dataset for 11 intuitive movement tasks from single upper extremity during multiple recording sessions,'' \emph{GigaScience}, vol.~9, no.~10, p. giaa098, 2020.

\bibitem{alwasiti2020motor}
H.~Alwasiti, M.~Z. Yusoff, and K.~Raza, ``Motor imagery classification for brain computer interface using deep metric learning,'' \emph{IEEE Access}, vol.~8, pp. 109\,949--109\,963, 2020.

\bibitem{van2008visualizing}
L.~Van~der Maaten and G.~Hinton, ``Visualizing data using t-sne.'' \emph{Journal of machine learning research}, vol.~9, no.~11, 2008.

\end{thebibliography}
\end{document}